\documentclass[11pt,twoside]{article}  
\usepackage{apn3conf}

\begin{document}   

%
%
%
%

\title{The Evolution of Nova Ejecta}
%

\author{Michael F.\ Bode}
\affil{Astrophysics Research Institute, Liverpool John Moores University, Birkenhead, CH41 1LD, UK. Email: mfb@astro.livjm.ac.uk}



\contact{Mike Bode}
\email{mfb@astro.livjm.ac.uk}

%
%

\paindex{Bode, M.F.}
%
%

\authormark{Bode}

%
%



\begin{abstract}          
In this paper I review the basic parameters of Classical Novae and then move on to describe the evolution of their ejected envelopes. The early shaping of the remnant, thought to be a consequence of a common envelope phase, and with analogies to what may occur in PNe with binary star nuclei, is then described. Finally, the curious case of Nova GK Persei (1901) and its potential to aid our understanding of both nova and long-term PN evolution is discussed.   
\end{abstract}


\section{Introduction}

Classical Novae (CNe) represent an important class of eruptive astronomical object with only GRB, Supernovae and some LBV's exceeding the energy released in their outbursts. The most recognisable characteristic is their visual light curve. Typically, this shows a rapid rise of less than a few days' duration followed by a slower decline. Indeed, the rate at which a nova declines away from peak defines its ``speed class'' with the slowest novae declining at less than 0.01 magnitudes per day and the fastest novae over ten times this rate. The speed class is in turn related to both the absolute magnitude at peak (in the sense that the faster the nova, the brighter it is intrinsically) and to the velocity of the principal ejecta (the faster the nova, the higher $v_{ej}$ -- see e.g. Warner 1989).

The parameters of the central binary systems of CNe are relatively well defined compared to those for symbiotic stars and suspected binary nuclei PNe (see e.g. Schwarz, Pollacco this volume). It has been known since the 1950's for example that CNe comprise a white dwarf and late-type main sequence star with binary period typically a few hours. The secondary (main sequence) star is losing matter onto the surface of the WD via an accretion disk. Every $10^{4} - 10^{5}$ years CNO cycle hydrogen burning occurs on the degenerate dwarf surface resulting in a thermonuclear runaway (TNR). This in turn leads to a rapid rise in luminosity ($\sim 10^{5}$ times) and ejection of $\sim 10^{-5} - 10^{-4}$M$_{\odot}$ at up to several thousand km s$^{-1}$. The nova outburst may recur several thousand times during the course of evolution of the system (see Warner 1995 and references therein). 

\section{Observations of the Expanding Ejecta}



Evidence that there is large scale order in the ejecta is seen in optical spectra at very early times after outburst. This has led in turn to models of the geometry of ejected material in terms of polar caps and tropical and equatorial rings for example (see Bode 2002 for a recent review). The earliest spatial resolution of nova shells has come from radio interferometry with the MERLIN array as little as two months after outburst (Eyres et al. 1996). There are however problems relating the observed structure in the radio to that deduced from optical spectroscopy (Eyres et al. 2000). This is at least in part due to our current inability to simultaneously observe at more than one frequency at this spatial resolution to disentangle the effects of temperature and optical depth on the radio images. The advent of e-MERLIN will solve this.   

\begin{figure}
\epsscale{.80}
\plotone{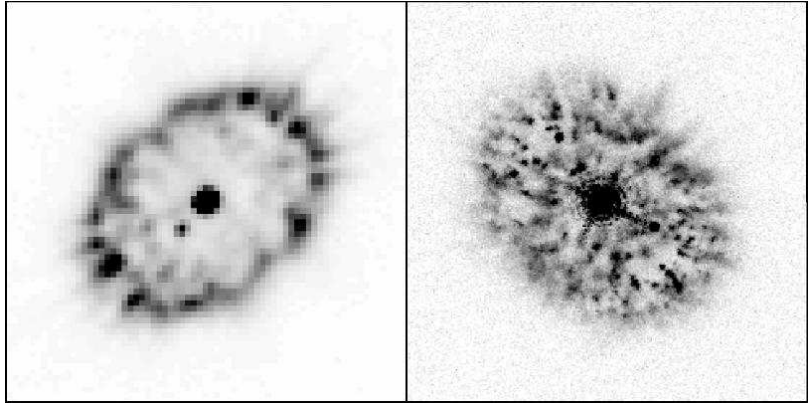}
\caption{H$\alpha$/[NII] images of DQ Her (with WHT, from Slavin, O'Brien
\& Dunlop 1995) and HR Del (HST, from Harman \& O'Brien 2003)} \label{T1.10-fig-1}
\end{figure}

To-date, optical imagery has resolved over 40 CNe remnants. This includes sevenresolved by HST (Bode 2002). Much of the recent work in this area has been undertaken by Tim O'Brien and his collaborators (e.g. Slavin, O'Brien \& Dunlop 1995; Gill \& O'Brien 1998, 1999, 2000). Figure~\ref{T1.10-fig-1} shows images of the moderate speed class nova DQ Her (1934) and the very slow nova HR Del (1967). One can clearly see from these images several pertinent features.

In both objects we observe ellipsoidal shells. At high spatial resolutions, the overall symmetry is seen to comprise many blobs and knots. The ejected shell of HR Del looks very different in H$\alpha$ than in [OIII] with the latter emission being concentrated to the poles of the remnant (Harman \& O'Brien 2003). In DQ Her, the remnant shows clear evidence of equatorial and tropical bands. In both remnants, tails of emission extend out from the main shell. This is suggestive of ablation of knots in the shell by a faster moving wind which follows some time after the initial eruption (Harman \& O'Brien 2003). In section 4 we show the image of the remnant of the very fast nova GK Per which may be compared with the remnants of the slower novae shown here.

The combination of imagery and long slit spectroscopy has allowed the determination of the precise geometry of several remnants (e.g. Gill \& O'Brien 2000). It has confirmed that the remnants are prolate ellipsoids and also allowed us to determine accurate distances to the remnants by expansion parallax (note that significant deceleration of the remnants is neither expected nor observed in the immediate decades following outburst, with the exception of GK Per which is discussed in detail below). Correction for the effect of inclination on the observed shape has also shown that there is a strong correlation between remnant shape and speed class in the sense that the slower the nova, the larger the ratio of major to minor axis dimensions (Bode 2002).

\section{Remnant Shaping}

For a typical nova, the ejecta from the WD envelop the secondary star in a matter of minutes. A common envelope binary thus results, with the secondary imparting energy and angular momentum to the ejecta. This scenario was first explored by in detail by Livio et al. (1990), then subsequently by Lloyd, O'Brien \& Bode (1997) and Porter, O'Brien \& Bode (1998).

Lloyd et al. modelled the evolution of the remnant using a 2.5D hydrodynamic code. The principal ejecta are represented by a wind with secularly increasing velocity but decreasing mass loss rate followed some time later by faster moving, more tenuous material. An important ingredient is the inclusion of the observed relation of $v_{ej}$ to speed class. For slower novae, lower ejection velocities effectively lead to a longer interaction time between the ejecta and the secondary and hence the expectation of a greater degree of shaping.

\begin{figure}
\epsscale{.80}
\plotone{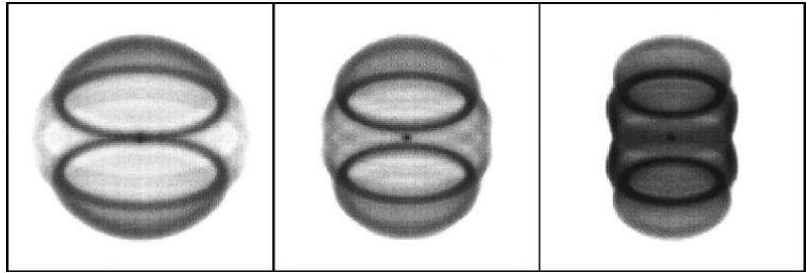}
\caption{Synthetic images resulting from hydrodynamical simulations of the evolution  of nova ejecta including WD envelope rotation (Porter et al. 1998). Left to Right, results are for 0, 0.5 and 0.7 of the Keplerian velocity at the WD surface for ejecta from a moderately fast nova.} \label{T1.10-fig-2}
\end{figure}

This model was successful in producing features such as rings and caps, plus a correlation of overall shape to speed class as required. However it gave rise to {\em oblate} shells. Porter et al. added modifications in terms of the inclusion of WD envelope rotation and its effects on the progress of the initial nova outburst with latitude on the WD surface. This again produced the required features of the remnants, but also {\em prolate} shells, as observed. Figure~\ref{T1.10-fig-2} shows some of the synthetic images produced by these latter simulations. Full 3D simulations are now in progress (see Wareing et al., this volume).

%
%

\section{Nova GK Persei 1901}


GK Per rivalled the brightness of Vega at the peak of its outburst in 1901 and then declined very rapidly. It is now classified as a very fast ``neon'' nova. The latter epithet comes from the strength of the Ne lines in its post-outburst spectrum, which in turn points to the eruption being on the surface of a high mass (ONeMg) WD. At a distance of 470pc (derived by expansion parallax for $v_{ej}=1200$ km s$^{-1}$) $L_{max}\sim 5\times10^{38}$ erg s$^{-1}$ ($\sim10^{4}$ times the quiescent luminosity) and $M_{ej}\sim10^{-4}$M$_{\odot}$ (see Seaquist et al. 1989 and references therein). Although the outburst itself is not unique among the observed CNe, its central system certainly is. 

The nova binary comprises a WD with a high value of surface magnetic field ($B\sim5\times10^{5}$ gauss --  an intermediate polar) together with an evolved late-type (K2IV) companion. The primary and secondary masses have been estimated as $0.9$M$_{\odot}$ and 0.25M$_{\odot}$ respectively (see Dougherty et al 1996 and references therein). The binary period is the longest known for a CN at 1.904 days (the long period being consistent of course with the existence of an evolved, Roche lobe-filling secondary in the system). Also unusual for a CN are dwarf nova-like outbursts which have been observed on several occasions since the main nova eruption of 1901.

In the months following the outburst, rapidly expanding nebulosities were seen on arcminute scales (Ritchey 1901). Despite the fact that the true distance to the nova was poorly determined at the time, the nebulosities were realised to be expanding at around or above light speed and were probably due to a geometrical effect (a ``light echo'' -- see also Bond, Sugarman this volume). Indeed, GK Per was the first astronomical source in which such an effect was observed.

\begin{figure}
\epsscale{.70}
\plotone{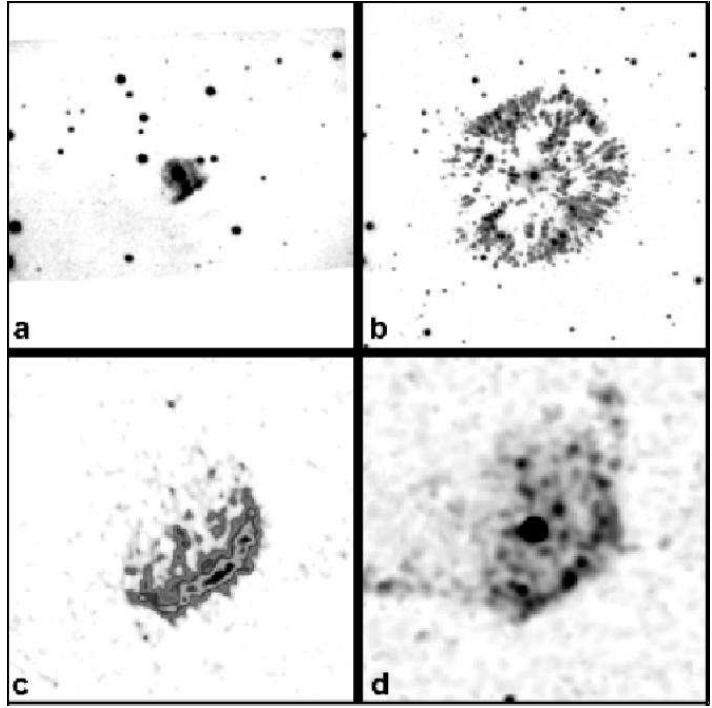}
\caption{ Images of the expanding ejecta of GK Per. (a) 1917, optical (Ritchey 1918); (b) 1993, H$\alpha$/[NII] (Slavin et al. 1995), (c) 1997, 5 GHz VLA (E.R. Seaquist, private communication) and (d) 2000, X-ray (0.4 - 1.0 keV, Balman 2002). All images are $3^{\prime}\times3^{\prime}$ on a side (N up, E left).} \label{T1.10-fig-3}
\end{figure}

Figures~\ref{T1.10-fig-3}a,b show the expanding ejecta in 1917 and 1993 respectively. The 1917 image appears to show the typical ring-like structure expected for the ejecta. However, superimposed to the SW is a bar of emission. Indeed, from following the sequence of later optical imagery, it appears that the ejecta are being decelerated by interaction with an unseen medium in this quadrant (Seaquist et al. 1989). In the 1993 image, the very blobby nature of the ejecta is evident.

The confirmation that a strong interaction is taking place in this direction comes from the detection of a ridge of non-thermal (synchrotron) radio emission, as shown in Figure~\ref{T1.10-fig-3}c. Furthermore, recent CHANDRA images (Figure~\ref{T1.10-fig-3}d) confirm that the region is a source of X-ray emission. The nova ejecta of GK Per now resemble a supernova remnant in miniature, but evolving on human timescales.

\begin{figure}
\epsscale{.70}
\plotone{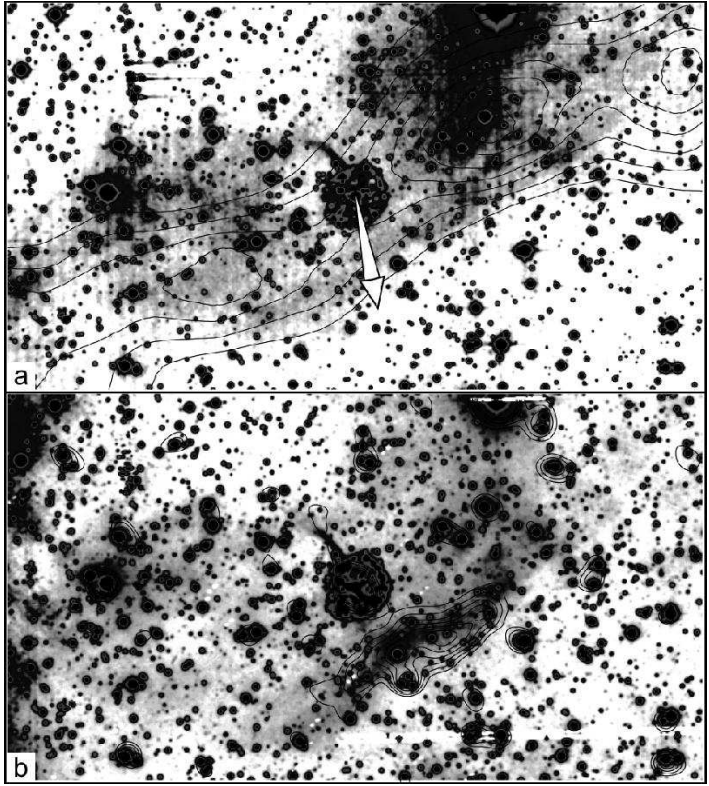}
\caption{Images of the extended nebulosity of GK Per. (a) [OIII]5007 filter with IRAS 100$\mu$m contours and (b) H$\alpha$ filter with contours of the light echo in 1902. The arrow shows the derived direction of proper motion of the central binary with 1$\sigma$ error bars. Images are $18^{\prime}\times11^{\prime}$.} \label{T1.10-fig-4}
\end{figure}

A search for the medium into which the 1901 nova ejecta are running resulted in the detection of a large scale ($\sim30^{\prime}$) nebula in the far infrared by IRAS (Bode et al. 1987). Subsequent re-analysis of HIRAS images by Dougherty et al. (1996 - see Figure~\ref{T1.10-fig-4}a) led to the confirmation that the extended far IR emission is due to dust at $T=23\pm1$K with $M_{d}=0.04$M$_{\odot}$ (compared to a mass of neutral hydrogen in the same region of around 1M$_{\odot}$).

Bode et al. (1987) first suggested that the IRAS emission might be due to material ejected in a previous phase of evolution of the central binary in the form of an ancient PN. Dougherty et al. later suggested that the origin of this material was a ``born again'' AGB star phase of the binary as the WD accreted material at a very high rate from the secondary star creating a common envelope which was then ejected. Dynamical considerations suggest that the age of this nebula is around 10$^{5}$ years, with the last major ejection around $3\times10^{4}$ years ago (for expansion velocities of 20 km s$^{-1}$). The Dougherty et al. evolutionary model is consistent with the current (low) secondary mass for its luminosity, and the $\sim1$M$_{\odot}$ of material present in the ancient nebula (which it would have lost).


Tweedy (1995) first detected extended optical nebulosities extending well beyond the current nova ejecta and correlated with some of the 1901/02 light echoes. We have since conducted more detailed and extensive observations using the Wide Field Camera of the 2.5m Isaac Newton Telescope on La Palma (Bode, O'Brien \& Simpson 2003). Two of the resulting images are shown in Figure~\ref{T1.10-fig-4}.

The extended emission is suggestive of a slightly tilted ``hourglass'' flattened towards the SW. The 1901 nova sits in the middle of the waist of the optical nebula and a saddle point of the IRAS emission (Figure~\ref{T1.10-fig-4}a). Careful inspection of the images shows that the [OIII] emission is interior to that in the H$\alpha$ filter, particularly towards the SW. The brightest region of the large scale optical nebulosity is also coincident with the longest lived of the light echoes from  1902 (Figure~\ref{T1.10-fig-4}b). The question to be answered is whether we can tie together the asymmetries seen in the nova and ancient nebulosities in some consistent way.

We suspected that the nova binary is moving relative to the ISM towards the SW. To address this, we took the 1917 and 1993 images and measured the proper motion of the central stars. The result is shown in Figure~\ref{T1.10-fig-4}a and confirms our expectation. In addition, together with measured radial velocities (Crampton, Cowley \& Fisher 1986), the space velocity of the system is found to be $45\pm4$km s$^{-1}$ (Bode et al. 2003).

We thus have a self-consistent model where the relatively high space velocity central binary went through a final born-again AGB phase of common envelope evolution around $10^{5}$ years ago and ejected $\sim1$M$_{\odot}$ of material in the form of a bipolar nebula. This nebula then interacted with the ISM in the direction of motion, suffering deceleration and thus giving rise to the rather flattened appearance we see now to the SW. The nova binary is however essentially a bullet that suffers no such deceleration and (as in the case of many elderly and ancient PN - see e.g. Kerber, this volume) ends up off-centre, in the direction of overall motion. 

In 1901, the central WD had accreted enough material from the secondary (this time at a slower rate) to undergo a nova outburst. The outburst illuminated the circumstellar and interstellar environment giving rise to the light echoes. The nova ejecta then encountered higher density material towards the SW, leading in turn to the shocks and particle acceleration (probably in the enhanced magnetic field drawn out from the WD) that we see now as X-ray and radio emission.

\section{Concluding Remarks}

The evolution of Classical Nova remnants is currently better understood than that of PNe as the parameters of the central system and ejecta in the former are generally more precisely defined. That being said, and acknowledging the fact that ejected masses, velocities and timescales are very different, it became apparent to me at this conference that the two communities could gain much from a closer working relationship. In the case of at least one nova (GK Per) we have an object of immediate and potentially significant common interest.

\end{document}